\documentclass[12pt,a4paper]{article}
\usepackage{graphicx}
\usepackage{amsmath}
\usepackage{mathtext}
\usepackage{amssymb}

\newcommand\beq{\begin{equation}}
\newcommand\eeq{\end{equation}}
\newcommand{\beqn}{\begin{eqnarray}}
\newcommand{\eeqn}{\end{eqnarray}}
\newcommand{\beqnn}{\begin{eqnarray*}}
\newcommand{\eeqnn}{\end{eqnarray*}}
\newcommand{\tr}{\mbox{Tr}}

\begin{document}
\sloppy 
~\vspace{-1cm}

\begin{flushright}
{ITEP-LAT/2005-28} 
\vspace{0.2cm} 
\end{flushright}
\begin{center}
{\baselineskip=24pt {\Large \bf On topological properties
of vacuum defects in lattice Yang-Mills theories}\\

\vspace{1cm}

{\large A.\,V.\,Kovalenko$^{\dag}$,
	S.\,M.\,Morozov$^{\dag}$, 
    M.\,I.\,Polikarpov$^{\dag}$ and
    V.\,I.\,Zakharov$^{*}$} } \vspace{.5cm} {\baselineskip=16pt { \it

$^{\dag}$ Institute of Theoretical and  Experimental Physics,
B.~Cheremushkinskaya~25, Moscow, 117259, Russia\\
$^{*}$ Max-Planck Institut f\"ur Physik, F\"ohringer Ring 6, 80805, M\"unchen,
Germany} }

\end{center}
\abstract{
We study correlations between low-lying modes of the overlap Dirac
operator and vacuum defects, center vortices and three-dimensional
volumes, in lattice SU(2) gluodynamics. The low-lying modes are
apparently sensitive to topological properties of the underlying gluon
field configurations while the vacuum defects are crucial for the
confinement. We find distinct positive correlation in both cases. In
case of vortices the correlation is stronger.
}
\date{}
\newpage

\section{Introduction}
We study properties of the low-lying modes of the Dirac operator
$$D_{\mu}\gamma_{\mu}\psi_n~=~\lambda_n\psi_n~~,$$
where the covariant derivative is constructed on the lattice vacuum
gluonic field configurations $\{A_{\mu}^a(x)\}$. Low-lying modes play a
special role in understanding topological properties of the gluonic
field configurations. Namely, the difference between number of zero
modes with positive and negative chirality is related to the total
topological charge of the lattice volume:
$$ n_{+}-n_{-}~=~Q_{top}~~.
$$
Near-zero mode modes determine the value of the quark condensate via
the Banks-Casher relation:
\begin{equation}
<\bar{q}q>~=~-\pi \rho(\lambda_n\to 0)~~,
\end{equation}
where  $\lambda_n\to 0$ with the total volume tending to infinity.
Properties of the low-lying fermionic modes were
studied in great detail for cooled gluon-filed configurations
and these studies confirmed the quasiclassical picture,
at least in its gross features, for review see, e.g., \cite{teper}.

Recently it turned possible to investigate the topological fermionic
modes working with the original fields $\{A_{\mu}^\alpha(x)\}$. The use
of the overlap operator~\cite{neuberger} is crucial for this purpose.
Measurement brought some unexpected results \cite{random}. Namely the volume occupied
by low-lying modes apparently tends to zero in the continuum limit of
vanishing lattice spacing, $a\to 0$,
\begin{equation}\label{vanishing}
\lim_{a\to 0}{V_{loc}}~\sim~a^{\alpha}~\to ~0~~,
\end{equation}
where $\alpha$ is a positive number of order unit (for details see
original papers \cite{random}) and the localization volume
(\ref{vanishing}) is defined in terms of the Inverse Participation
Ratio (IPR).

Observation (\ref{vanishing}) implies that there exists short-distance
description of the topological properties of the vacuum. In particular,
one can ask, where the modes shrink to.

Independently of the measurements on topological fermionic modes, there
accumulated evidence that lower-dimensional vacuum defects are crucial
for the confinement. We have in mind center vortices and 3d volumes.
The role of the center vortices has been discussed since long, for
review see \cite{greensite}. These defects are defined as 2d surfaces.
However, commonly one was assuming that these, by definition thin
vortices are void of physical meaning and only mark, approximately
`thick' vortices. Measurements reported in Ref \cite{ft} indicate,
however, that thin vortices possess remarkable gauge invariant
properties, for review see, e.g., \cite{vz} The role of
three-dimensional defects \cite{3d} is highlighted by observation
\cite{forcrand} that by changing a part of lattice in a special way
(defined in terms of projected fields) one loses confinement and
nullifies the quark condensate. This part of the lattice turns to be,
up to gauge transformations, a three-dimensional volume \cite{3d}.

In this note we report results of measurements of correlation between
low-lying fermionic modes and vacuum defects.

\section{Measurements}
\subsection{Definitions}

We use standard definitions of the center vortices in the Direct
Maximal Center Projection~\cite{greens2} which is defined in SU(2)
lattice gauge theory by the maximization of the functional
\begin{equation}
F(U) = \sum_{n,\mu} \left( \tr U_{n,\mu}\right)^2 \, , \label{maxfunc}
\end{equation}
with respect to gauge transformations, $U_{n,\mu}$ is the lattice gauge
field. The maximization of (\ref{maxfunc}) fixes the gauge up to Z(2) gauge
transformations and the corresponding Z(2) gauge field is defined as:
$Z_{n,\mu} = \mbox{sign} \tr U_{n,\mu}$. The plaquettes $Z_{n,\mu\nu}$
constructed as product of links $Z_{n,\mu}$ along the border of the plaquette
have values $\pm 1$. The P-vortices (forming closed surfaces in 4D space) are
made from the plaquettes, dual to plaquettes with $Z_{n,\mu\nu} = -1$.

We have computed the eigenvalues $\lambda_{lat}$ and the eigenfunctions $\psi_\lambda(x)$
of the overlap Dirac operator on the lattice with $\beta = 2.45$ and
$L_s = L_t = 14$ by solving an eigenvalue problem
\beqn
D_{ov} \psi_\lambda(x) = \lambda_{lat} \psi_\lambda(x).
\eeqn

To make a connection with continuum physical eigenvalues $\lambda$ we
have stereographically project $\lambda_{lat}$ onto the imaginary axis
and divided by lattice spacing~\cite{zenkin},
\beqn
\lambda = Im\left(\frac{\lambda_{lat}}{1-\lambda_{lat}/2}\right) /a.
\eeqn

The scalar density of eigenmode is defined by
\beqn
\rho_\lambda(x) = \psi^\dagger_\lambda(x)\psi_\lambda(x), \ \
\sum_x\rho_\lambda(x) = 1\, .
\eeqn

\subsection{Correlation between vortices and fermionic modes}

To clarify the role of the vortices in the topological structure of the
vacuum we measure the correlator between intensity of fermionic modes
and of center vortices. The correlator depends on the eigenvalue and on
the local geometry of the vortex. The vortex lives on the dual lattice
and we consider the correlator of the points on the dual lattice,
$P_i$, which belong to center vortex with the value of $\rho(x)$
averaged over the vertices of the 4d hypercube, $H$, dual to $P_i$.
Thus we consider the correlator:
\begin{equation}
C_\lambda = \frac{\sum_{P_i} \sum_{x \in H} ( V \rho_\lambda(x) -
\langle V\rho_\lambda(x)\rangle)} {\sum_{P_i} \sum_{x \in H}1} \, .
\label{eq:z2_rho_corr}
\end{equation}
\begin{figure}
\centerline{\includegraphics{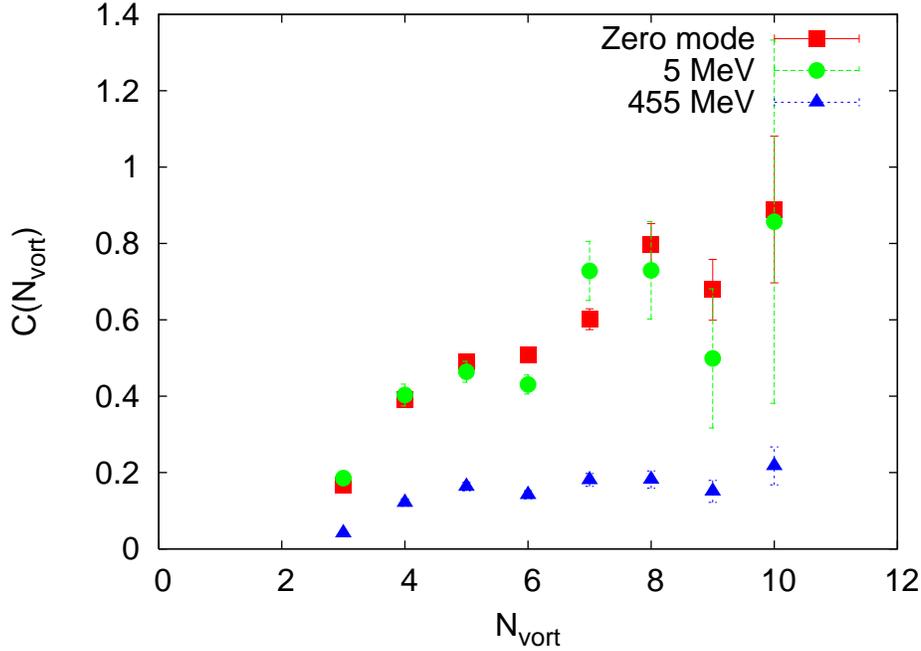}}
\caption{Dependence of $C_\lambda$~(\ref{eq:z2_rho_corr}) on $N_{vort}$
for $\beta=2.45$, $L^4=14^4$.} \label{Clambda}
\end{figure}
We find out  that this correlator strongly depends on the number of the
vortex plaquettes, $N_{vort}$ attached to point $P_i$. Numerical
results for $C_\lambda$ are shown on Fig.~\ref{Clambda}. We see that
the more plaquettes are attached to given point, the larger correlator
is. Also, the larger eigenvalue $\lambda$ is, the smaller is the correlator.
These results are in agreement with the general picture that vortices
are related to chiral symmetry breaking, which is due to the low lying
eigenmodes.

\subsection{Correlation between three-dimensional volumes and Dirac eigenmodes}

\begin{figure}
\centerline{\includegraphics{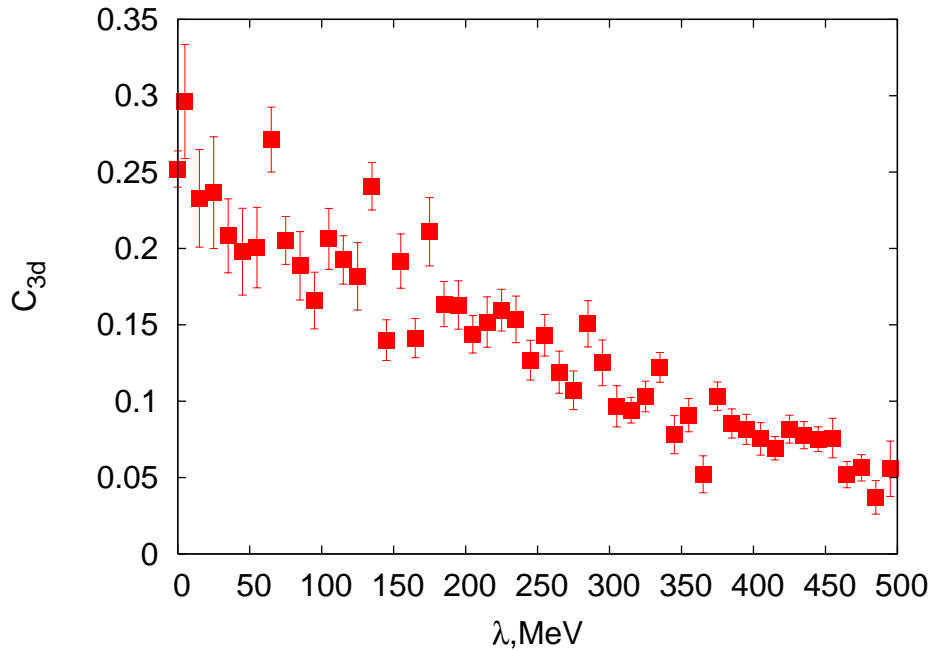}}
\caption{Dependence of the correlator (\ref{Co3d}) on the energy of the
Dirac eigenmode, $\lambda$.} \label{C3d}
\end{figure}

By gauge transformation we can minimize the number of negative links in
$Z(2)$ projection from which we construct center vortices. This gauge
is called Z(2) Landau gauge. These links are dual to 3d cubes on the
dual lattice, these cubes form 3d volumes, which scale in physical
units~\cite{3d}. Thus these negative links (or corresponding 3d
volumes) play the role in the confinement. Removing these negative
links leads to gauge field configurations with zero string tension. We
calculated in Z(2) Landau gauge the correlator
\begin{equation}\label{Co3d}
C_{3d} = \frac{(\sum_s V\rho_\lambda - <V\rho_\lambda>)}{\sum_s 1}.
\end{equation}
where the sums are over sites, $s$, which are  endpoints of negative
links. This correlator can be considered as the correlator of dual 3d
volumes and Dirac eigenmode with the energy $\lambda$. The result is
shown on Fig.~\ref{C3d}. The larger eigenvalue is, the smaller
correlation we observe, from general grounds it is clear that the high
energy eigenmodes should not be related to topology and confinement.

\section{Conclusions}

We have found strong evidence for correlation between local vortices
density and density of low-lying eigenmodes of the overlap Dirac
operator. The correlation of fermionic modes with density of
three-dimensional volume is also positive, although weaker. These
observations support strongly the idea that there exists picture of
confining fields and of topological structures in the vacuum state of
YM theories in terms of short distances.

\section*{Acknowledgments}
Numerous discussions with F.V.~Gubarev on the topics of this paper are
gratefully acknowledged. The invaluable assistance of G.~Schierholz and
T.~Streuer in the overlap operator implementation is kindly
acknowledged. This work was partially supported by grants
RFBR-05-02-16306a, RFBR-0402-16079 and  EU Integrated Infrastructure
Initiative Hadron Physics (I3HP) under contract RII3-CT-2004-506078
(MIP);RFBR-05-02-16306a and RFBR-05-02-17642 (S.M.M. and A.V.K.). M.I.P. is
happy to acknowledge the hospitality of the theory group of MPI
(M\"unchen), where part of this work was carried out.


\end{document}